\documentclass[preprint,aps]{revtex4}
\begin{document}
\title{Compatibility conditions on local and global spectra for 
$n$-mode Gaussian states}
\author{J. Solomon Ivan}
\email{solomon@imsc.res.in}
\affiliation{The Institute of Mathematical Sciences,  CIT Campus, Taramani, 
Chennai 600 113, India}
\author{R.  Simon}
\email{simon@imsc.res.in} 
\affiliation{The Institute of Mathematical Sciences,  CIT Campus, Taramani, 
Chennai 600 113, India}

\date{November 15, 2008} 

\begin{abstract}
Compatibility conditions between the (global) spectrum of an 
$n$-mode Gaussian state and the spectra of the individual modes 
are presented, making optimal use of 
beam splitter and (two-mode) squeezing transformations.  
An unexpected bye-product of our elementary approach is 
the result that 
every two-mode Gaussian state is uniquely determined, modulo 
local transformations, by its global spectrum and  local 
spectra -- a property shared not even by a pair of qubits.
\end{abstract}
\maketitle

The quantum marginal problem has attracted considerable interest in 
quantum information 
theory\,\cite{higuchi,bravyi,klyachko,han,christandl,hall,liu,nielsen,hiro}. 
Given a multipartite system, it asks:  
what kind of spectra for the subsystem density operators  
are consistent with a given 
spectrum for the density operator of the full system? 
The Gaussian quantum marginal problem (detailed below)  
 has been solved 
recently\,\cite{hiroshima,eisert}  
(As noted in Ref.\,\cite{eisert},  
the three-mode case was known earlier\,\cite{adesso}).  
 Our approach to this problem  makes effective use of beam splitter 
and two-mode squeezing transformations.  In the  
case of two modes 
it is shown that {\em every  
 Gaussian state is uniquely determined, modulo 
local canonical transformations, by its global spectrum and  
local spectra}; {\em in particular, the entanglement   
 is  fully determined by these spectra}.

Consider a Gaussian state of a system of $n$-modes, 
represented by density 
operator $\hat{\rho}$. The mean values of the position and momentum 
variables
$q_j,\;p_j$ have no role to play in our considerations,  and
 so we assume that these mean values vanish.
Such a zero-mean Gaussian state is fully described by its $2n \times 2n$ 
covariance matrix $V$.

The reduced state $\hat{\rho}_{j}$ of the $j^{\rm th}$ mode, obtained by tracing
out from $\hat{\rho}$ all other modes, is also a zero-mean Gaussian state. With
the phase space variables assumed arranged in the order $q_1,p_1\,;\,q_2,p_2\,
;\,\cdots\,;\,q_n,p_n$ the $j^{\rm th}$ $2 \times 2$ block along the 
leading diagonal of $V$ 
represents precisely the covariance matrix of the reduced state $\hat{\rho}_j$.
Through (independent) local canonical transformations $\in Sp(2,R)$ on 
each mode we make
all the $2 \times 2$ blocks along the diagonal of $V$ multiples of identity.
The covariance matrix of the $j^{\rm th}$ mode will then be of the form 
diag$(m_j,m_j)$. It corresponds to a thermal state, with  
temperature $T(m_j)$ which is a monotone increasing function of $m_j$. 
 Being thermal, 
$\hat{\rho}_j$  has the spectral resolution
$\hat{\rho}_j=[1-\xi(m_j)]\sum_{k=0}^{\infty}{\xi(m_j)}^{n_{jk}} |n_{jk} 
\rangle \langle n_{jk}|\,.\,$
The parameter $\xi(m_j)$ is another monotone increasing function of 
$m_j$, and 
$|n_{jk} \rangle$'s 
are the energy eigenstates  of the $j^{\rm th}$ oscillator. 
 Clearly, 
the eigenvalue spectra of the $\hat{\rho}_j$'s are determined by, 
and 
determine, the {\em local spectral parameters} $m_j$. 

Using an appropriate (nonlocal) canonical transformation $S \in Sp(2n,R)$ 
the 
covariance matrix $V$ can be decoupled and brought into the canonical
form $V^{(0)}$ of independent oscillators in thermal states\,\cite{simon}: 
$V^{(0)}=SVS^{T}$
=${\rm diag}(\kappa_1, \kappa_1\,;\,\kappa_2,\kappa_2\,;\,\cdots\,;\,
\kappa_n, \kappa_n)$. The associated density operator $\hat{\rho}^{(0)}$
thus has the spectral decomposition
\begin{eqnarray}
\hat{\rho}^{(0)}= \prod_{j=1}^{n}[1-\xi(\kappa_j)]\sum_{k=0}^{\infty}
{\xi(\kappa_j)}^{n_{jk}} | n_{jk} \rangle \langle n_{jk} |.
\end{eqnarray}
Since $\hat{\rho}^{(0)}$ and the original $\hat{\rho}$ are unitarily
related, the  spectrum of $\hat{\rho}$ is the same as that of 
$\hat{\rho}^{(0)}$. It is clear that this global spectrum and the 
$n$-tuple of {\em global spectral 
parameters} 
  $(\kappa_1,\kappa_2,\cdots,\kappa_n)$ determine each other.

We may now ask what are the
constraints connecting the global spectrum of a Gaussian state to its local
spectra. In view of the invertible relationships just noted 
  this {\em Gaussian quantum marginal problem} is equivalent to 
seeking the compatibility constraints between
the global spectral parameters $\{\,\kappa_j\,\}$ and the 
  local spectral parameters $\{\,m_j\,\}$. Interestingly, the answer can 
be given in the form of necessary and sufficient conditions.

{\em Theorem}: Let $m=(m_1, m_2, m_3, \cdots, m_n)$ and $\kappa= (\kappa_1,
\kappa_2, \cdots, \kappa_n)$ be the local and global spectral parameters 
 of an $n$-mode Gaussian state, written  in
nondecreasing order. These are  compatible iff
\begin{eqnarray}
&&\sum_{j=1}^{k} m_j \geq \sum_{j=1}^{k} \kappa_j, \,\,\,\,k=1,2,\cdots, 
n\,,\\
&&m_n- \sum_{j=1}^{n-1}m_j \leq \kappa_n -\sum_{j=1}^{n-1} \kappa_j.
\end{eqnarray} 

{\em Remarks}: What this claim means can be clarified by stating it 
in two 
parts. Suppose a Gaussian state 
is given. Its local spectral parameters $m_1, m_2, \cdots, m_n$, and
global spectral parameters $\kappa_1, \kappa_2, \cdots, \kappa_n$ are
certain to meet these inequalities (with $\kappa_1 \geq1$). Conversely, 
given a set of local and global spectral parameters meeting these inequalities
(with $\kappa_1 \geq1$), we can certainly construct a physical Gaussian 
state with these parameters. 

The first part of the theorem was essentially proved by 
Hiroshima\,\cite{hiroshima}. But the full theorem in this form 
 was  
formulated by Eisert et al.\,\cite{eisert} who  presented an 
inductive proof for the second part. Our proof of both parts 
 will be seen to be constructive, consistent with the elementary 
nature of the theorem, and {\em it rests in an essential manner on a 
fuller appreciation of the two-mode situation}.

Given two vectors $m,\,\kappa\in R^n$, we will say  $\kappa$ {\em 
dominates} $m$ if $m$ and $\kappa$, {\em after their components are 
 rearranged 
in the nondecreasing order}, obey the set of $n+1$ inequalities 
(2), (3). 
 This definition is such that {\em permutation of the components
 of $m$ or $\kappa$ does not 
affect dominance}.  Thus $(9,7,8,6,12,11,10)$ is dominated by  
$(5,2,18,4,1,12,3)$, since $(1,2,3,4,5,12,18)$
 manifestly dominates $(6,7,8,9,10,11,12)$. 
 Further, dominance so defined is transitive: 
$\kappa$ dominates $m$, and $m$ dominates $m'$, together imply   
$\kappa$ dominates $m'$.  

In the Schur-Horn case \,\cite{horn} wherein $m$ corresponds to the 
diagonal 
entries of a hermitian matrix and $\kappa$ to its eigenvalues, 
the last inequality in (2) becomes an equality. It is clear  
that (3) is subsumed by (2) in that case. 

{\em The case of two modes}: 
This case is of interest in its own right. 
Further, it 
possesses an aspect which seems to be unique, not shared by any other system.
Finally, our analysis of the $n$-mode case relies critically on  
repeated 
applications of the two-mode result. Hence we begin with a 
direct proof of the theorem in the two-mode case. 

{\em Lemma\,}: The parameters $m_1 \leq m_2$ and 
  $1 \leq \kappa_1 \leq \kappa_2$ are compatible  for
two-mode Gaussian states iff 
\begin{eqnarray}
&&m_1 + m_2 \geq \kappa_1 + \kappa_2, \nonumber \\
&& m_2 - m_1 \leq \kappa_2 - \kappa_1.
\end{eqnarray} 
Note that the condition $m_1 \geq \kappa_1$ is subsumed by (4). 

{\em Proof of Lemma}: The covariance matrix can, through local unitary 
(canonical) transformation $\in Sp(2,R)\times Sp(2,R)$,
 be brought to the form
\begin{eqnarray}
V = \left(
\begin{array}{cccc}
m_1 & 0 & k_x & 0 \\
0 & m_1 & 0 & k_p \\
k_x & 0 & m_2 & 0 \\
0 & k_p &0 & m_2
\end{array}
\right).
\end{eqnarray}
The global spectral parameters $\kappa_1$, $\kappa_2$ are related to the 
local $m_1$, $m_2$ through the symplectic invariants\,\cite{simon} 
\begin{eqnarray}
\frac{1}{2} {\rm tr}\,(\Omega V \Omega^{T} V) &=& 
{\kappa}_{1}^{2} + {\kappa}_{2}^{2} ={m}_{1}^{2} + {m}_{2}^{2} + 
2k_x k_p, \nonumber \\
{\rm det}\,V = (\kappa_1\kappa_2)^2 &=& 
(m_1m_2-{k}_{x}^{2})(m_1m_2-{k}_{p}^{2}).
\end{eqnarray}
These  immediately imply
\begin{eqnarray}
\kappa_1 \kappa_2 &\leq& m_1 m_2, \nonumber \\
{\kappa}_{1}^{2} + {\kappa}_{2}^{2} &\geq& {m}_{1}^{2} + {m}_{2}^{2}, \;\; 
{\rm if}\,\,\,k_x k_p \geq 0, \nonumber \\
{\kappa}_{1}^{2} + {\kappa}_{2}^{2} &\leq& {m}_{1}^{2} + {m}_{2}^{2}, 
\;\; {\rm if}\,\,\,k_x k_p \leq 0,
\end{eqnarray}
equality in the first inequality holding if $k_x=0=k_p$. 
These inequalities imply 
\begin{eqnarray}
\kappa_2 - \kappa_1 &\geq& m_2 - m_1, \,\,\,{\rm when}\,\,\, k_x k_p \geq 0,
\nonumber \\
\kappa_2 + \kappa_1 &\leq& m_2 + m_1, \,\,\, {\rm when}\,\,\, k_x k_p \leq 0.
\end{eqnarray}
This much is immediate from the symplectic invariants. What  
remain to be proved are : $\kappa_2 - \kappa_1 \geq m_2 - m_1$ when
$k_x k_p \leq 0$ and $\kappa_2 + \kappa_1 \geq m_2 + m_1$ when 
$k_x k_p \geq 0$.

To prove these we reinterpret (6) as simultaneous expressions for
 $k_x$, $k_p$ in terms of $\kappa_1,\, \kappa_2;\,  m_1,\, m_2$:
\begin{eqnarray}
k_x k_p &=& [\,({\kappa}_{1}^{2} + {\kappa}_{2}^{2}) - ({m}_{1}^{2} + 
{m}_{2}^{2})\,]/2, \\
{k}_{x}^{2} + {k}_{p}^{2} &=& \frac{1}{m_1m_2} [\,{m}_{1}^{2} {m}_{2}^{2} 
- 
{\kappa}_{1}^{2} {\kappa}_{2}^{2} + {k}_{x}^{2} {k}_{p}^{2}\,].
\end{eqnarray}
It is clear that real solutions for $k_x$ and $k_p$ will exist iff  
`$\,{k}_{x}^{2} + {k}_{p}^{2}\,$'  $\geq$ `$\,2|\,{k}_{x} {k}_{p}\,|\,$'. 
That is, iff 
\begin{eqnarray}
m_1 m_2 - |\,{k}_{x}^{} {k}_{p}^{}\,| \geq \kappa_1 \kappa_2.
\end{eqnarray}
With use of (9) for $k_x k_p$, this last condition reads 
\begin{eqnarray}
\kappa_2 - \kappa_1 &\geq& m_2 - m_1, \,\,\,{\rm when}\,\,\, k_x k_p \leq 0,
\nonumber \\
\kappa_2 + \kappa_1 &\leq& m_2 + m_1, \,\,\, {\rm when}\,\,\, k_x k_p \geq 
0.
\end{eqnarray}
Proof of the Lemma is thus complete.

Two types of simple transformations on any pair of modes characterised by 
annihilation operators $a_j,\,a_k$ deserve particular mention; they play 
a key  role in our proof of the theorem. The first,  
$S_\theta$, corresponds to the {\em compact} transformations 
$a_j \to \cos \theta\, a_j + \sin \theta\, a_k,\,\,  
a_k \to -\sin \theta \,a_j + \cos \theta \,a_k$, and therefore is 
represented by 
$S_\theta = \cos \theta \,\sigma_0 \otimes \sigma_0 + \sin \theta\, 
i\sigma_2 \otimes \sigma_0 \in Sp(4,R),\; 0\le\theta<2\pi$, 
where $\sigma_0$ is the   $2\times 2$ unit matrix and $\sigma_2$ is the 
antisymmetric  Pauli matrix. Physically, $S_\theta$ is a {\em beam 
splitter with transmitivity} $\cos ^2\theta$. 
The second one,  $S_\mu$, 
is {\em noncompact} and corresponds to  squeezing transformations      
 $a_j \to \cosh \mu \, a_j + \sinh \mu \,a_k^\dagger,\,\,  
a_k \to \cosh \mu a_k + \sinh \mu\, a_j^\dagger$, 
and  is represented by 
$S_\mu = \cosh \mu \,\sigma_0 \otimes \sigma_0 + \sinh \mu \,\sigma_1
\otimes \sigma_3 \in Sp(4,R),\; 0\le\mu<\infty$.  

It is easily verified that when the  
covariance matrix $V$, Eq.\,(5), has $k_p=k_x\equiv k$, it can be 
diagonalized by the beam splitter transformation 
$V\to S_\theta V S_\theta^{\,T}$, with $\theta$ fixed through 
$\tan 2\theta = 2k/(m_2-m_1)$. And $\kappa_2 +\kappa_1$ will precisely 
equal $m_2+m_1$ in this case. Similarly, if 
$k_p= -k_x = k>0$, then $V$ is diagonalized by the squeezing 
transformation  $V\to S_\mu V S_\mu^{\,T}$, with  
$\tanh 2\mu = 2k/(m_2+ m_1)$, and one will find 
 $\kappa_2 -\kappa_1 = m_2-m_1$ in this case. 

Conversely, suppose we start with the canonical form 
$V^{(0)} = {\rm diag}\, (\kappa_1,\kappa_1;\kappa_2,\kappa_2)$, and we 
wish to achieve through symplectic congruence 
$V^{(0)}\to S V^{(0)}S^T, \;S\in Sp(4,R)$, a covariance matrix with 
diagonals $m_1,m_2$. If $m_1<m_2$ are such that $m_2<\kappa_2$ and 
 $\kappa_2 +\kappa_1 = m_2+m_1$, such a {\em redistribution} of  
 $\kappa_1,\,\kappa_2$ among  $m_1,\,m_2$ can always be achieved through a 
beam splitter transformation $S_\theta$.  
    Under $S_\theta$  we have $m_2 +m_1 = \kappa_2 +\kappa_1$ and 
 $m_2 - m_1 =\cos\,2\theta\,( \kappa_2 - \kappa_1)$.
  On the other hand, if $m_2>\kappa_2$ and 
 $\kappa_2 - \kappa_1 = m_2- m_1$, so that 
 $\kappa_1$ and $\kappa_2$ are enhanced by 
equal amounts to $m_1,\,m_2$, this  can  
 be achieved through a squeezing  transformation $S_\mu$.  
   Under $S_\mu$  we have $m_2 -m_1 = \kappa_2 -\kappa_1$ and 
 $m_2 +m_1 =\cosh\,2\mu\,( \kappa_2 +\kappa_1)$.

Our Lemma is similar to Lemma~5 of Ref.\,\cite{eisert}, but our proof 
is direct and constructive. There is an important distinction in content 
as well: while theirs claims that 
  $m_2 -m_1 = \kappa_2 -\kappa_1$ iff 
   $m_2 = \kappa_2$ and  $m_1 = \kappa_1$, 
 we have just demonstrated that if $m_2 -m_1 = \kappa_2 -\kappa_1$ then
   $m_2 + m_1$ could equal $\cosh\,2\mu\, (\kappa_2 +\kappa_1)$ 
for any $0\le\mu<\infty$, not just $\mu = 0$. Indeed, this 
distinction is central to Stage~2 of our proof of the second part of the 
 main theorem, the part which distinguishes the present symplectic 
situation from  the Schur-Horn case.

Returning to Eq.\,(10), if we are given values for the expressions  
`$\,a^2 + b^2\,$' and `$\,ab\,$' with $a^2 + b^2 \geq 2|\,ab\,|$, 
the solution for $(a,b)$ is {\em unique} [\,$(a,b)$ and $(b,a)$ are not 
distinct 
solutions for our purpose\,]. This innocent looking observation leads to a
surprising conclusion. 

{\em Proposition\,}: Specification of the local and global 
spectra of a two-mode
Gaussian state  determines {\em uniquely} the state itself, modulo local 
unitary transformations.

States of a pair of qubits share a similarity with two-mode Gaussian states in
 important respects. For instance,  positivity under partial transpose is 
a necessary and 
sufficient condition for separability and nondistillability in both cases.
But a statement analogous to the above proposition is not true for 
 a pair of qubits! 

{\em Proof of main theorem}: Assume we are given a (zero-mean) 
Gaussian state, or
equivalently, an acceptable covariance matrix $V$,  the 
$2 \times 2$ blocks along the leading diagonal of $V$ being of the form 
${\rm diag}(m_j, m_j)$. The global spectral parameters $\{\,\kappa_j\,\}$ 
are immediately
defined by $V$\,\cite{simon}. It is assumed that $m=(m_1,m_2, \cdots 
m_n)$ and 
$\kappa = (\kappa_1, 
\kappa_2,
\cdots, \kappa_n)$ are arranged in   nondecreasing order. Let 
$P_{\kappa}$
denote the product $\kappa_1 \kappa_2 \cdots \kappa_n$ and let $P_{m}=m_1 m_2
\cdots m_n$. Clearly,  $P_{\kappa} = {\rm det}\, V \leq P_m$, 
equality
holding iff  $V$ is  diagonal, i.e., iff 
$m_j = \kappa_j$, $j=1,2, \cdots, n$. Our task is to prove that 
$\kappa$
dominates $m$.

Choose a pair $1\leq j < k \leq n$ such that the $2 \times 2$ block (in 
the off-diagonal location)  
connecting the $j^{\rm th}$ and $k^{\rm th}$ modes is nonzero. We can 
arrange (through local rotations) 
this block to be diagonal. Let us `diagonalize' this $4 \times 4$ part of
the covariance matrix using an appropriate two-mode canonical transformation
$\in Sp(4, R)$, so that $m_j$ and $m_k$ are transformed to 
$\tilde{m}_{j}$ 
and $\tilde{m}_k$
respectively, the other diagonal parameters remaining unaffected.

It is be noted that the new $m$ dominates the original $m$. That this 
is so
follows, in the case $k < n$, from the facts $\tilde{m}_j < m_j$ and 
$\tilde{m}_j + \tilde{m}_k \le m_j + m_k$. In the case $k=n$ it follows 
from 
the 
additional
fact that if $\tilde{m}_k$ is less that $m_k$  it is so by a 
magnitude  which does  
not exceed the magnitude by which $\tilde{m}_j$ is less than $m_j$ 
$(\tilde{m}_k -\tilde{m}_j \geq m_k - m_j)$. 
Further, $\tilde{m}_j\tilde{m}_k < m_jm_k$.

  Denote by $m'$ the new diagonal $m$-parameters
arranged in nondecreasing order by correspondingly permuting the 
oscillators.   
 Since $\tilde{m}_j\tilde{m}_k < m_jm_k$ we have $P_{m'} < P_m$.

For purpose of  clarity, let us carry out this process   
 one more time.
The  parameters $m'$ will then go to $m''$ dominating 
$m'$,  with
$P_{m''} < P_{m'}$. It follows from the transmitivity 
of dominance that $m''$ dominates $m$.

It is now clear that when this process is iterated,  $m$ goes through a 
sequence of intermediate values, the value at every stage dominating the 
previous value, and correspondingly $P_m$ steadily decreasing, until
$P_m$ reaches $P_\kappa$ or, equivalently, until $V$ becomes 
diagonal. 
This completes 
proof of the first part of the theorem.
 
The elementary nature of our proof may be compared with that of 
Ref.\,\cite{hiroshima}. 
  $P_m$  played  
the role of `profit function' monitoring progress of this 
diagonalization process.

To prove the second part assume, conversely,  that we are given 
the global and local spectral parameters
$\kappa, \,m \in R^n$.
 Assume that these 
are compatible: i.e., $\kappa$ dominates $m$, with 
$\kappa_1 \geq 1$. Our task is to construct a Gaussian state with these 
properties. In other words we have to present a canonical transformation
$S \in Sp(2n,R)$ which acting on a covariance matrix  
 $V={\rm diag}(\kappa_1, \kappa_1 \,;\, \kappa_2, \kappa_2 \,;\, \cdots
\, ;\, \kappa_n, \kappa_n)$ will produce a covariance matrix $SVS^T$ with the 
target diagonal values $m$. We  build such an $S$ as a product
of $n-1$ {\em specific} two-mode transformations, evolving 
$m^{(0)}\equiv \kappa$  successively through a
sequence of intermediates $m^{(1)}$, $m^{(2)}$, $\cdots\,$ to finally 
$m^{(n-1)} 
= m$. It will be manifestly clear  
that $m^{(k)}$ dominates $m^{(k+1)}$ at each stage. For clarity,
this process is implemented through four elementary stages. 

{\em Stage 1}: Since $m^{(0)} \equiv \kappa$ dominates $m$, we have $m_1 
\geq 
{m}_{1}^{(0)}= \kappa_1$. Suppose $m_1 = {m}_{1}^{(0)}+\epsilon_1$, $\epsilon_1
>0$ (one will  move to the next step if $m_1 = {m}_{1}^{(0)}$).
Let $j_1$ be the least integer $<n$ such that ${m}_{j_1}^{(0)} \geq m_1$. 
Carry
out a beam splitter transformation $S_\theta$ between the first and   
${{j}_{1}}^{\rm 
th}$
mode so that the corresponding diagonal elements $({m}_{1}^{(0)}, 
{m}_{j_1}^{(0)})$
get redistributed to $({m}_{1}^{(0)} + \epsilon_1, {m}_{j_i}^{(0)}- 
\epsilon_1)=
({m}_{1}, {m}_{j_1}^{(0)}-\epsilon_1)$, with no change in the other 
diagonal 
entries:
$m^{(0)} =  ({m}_{1}^{(0)},{m}_{2}^{(0)}, \cdots, {m}_{n}^{(0)}) 
\rightarrow 
m^{(1)}  =  ({m}_{1},{m}_{2}^{(0)},\cdots,{m}_{j_1}^{(0)}-\epsilon_1, 
 \cdots,{m}_{n}^{(0)} ) 
\equiv ({m}_{1}^{},{m}_{2}^{(1)},m_3^{(1)},\cdots,{m}_{n}^{(1)} )$.
   
We can repeat this process. Let $m_2={m}_{2}^{(1)} + \epsilon_2$. By hypothesis
$\epsilon_2 \geq 0$ (this is so even if $j_1$ had equalled 2). Assume
$\epsilon_2 > 0$ (if $\epsilon_2 =0$, one moves to the next step). Let
$j_2$ be the smallest integer $<n$ such that ${m}_{j_2}^{(1)} \geq m_2$  
[Clearly,
$j_2$ can be as small as $j_1$, but not any smaller]. Carry out a 
beam splitter
transformation on the $2^{\rm nd}$ and  ${j}_{2}^{\rm th}$ modes 
so that the corresponding diagonal elements $({m}_{2}^{(1)}, {m}_{j_2}^{(1)})$
get redistributed to $({m}_{2}, {m}_{j_2}^{(1)} - \epsilon_2)$ to produce
$m^{(2)}$, leaving  the other diagonals unaffected.

If we are able to repeat this process only $\ell$ times we have, at the 
end of it,
\begin{eqnarray}
m^{(\ell)}=(\,m_1,m_2, \cdots, m_\ell; {m}_{\ell+1}^{(\ell)}, 
{m}_{\ell+2}^{(\ell)}, 
\cdots, 
{m}_{n}^{(\ell)}\,),
\end{eqnarray}
with ${m}^{(\ell)}_{j} < m_j$, $\forall$ $\ell+1 \leq j \leq n-1$, and 
${m}_{n}^{(l)}
={m}_{n}^{(0)}=\kappa_n$. 
What we have done so far is identical to what one would have done in the
Schur-Horn situation. Clearly, the beam splitter transformations carried
out so far affected neither the sum of the diagonal entries of 
$m^{(\cdot)}$ nor its $n^{\rm th}$ entry.
Consequently, the difference ${m}_{n}^{(k)} -
\sum_{j=1}^{n-1} {m}_{j}^{(k)}$
has remained the same for all $0 \leq k \leq \ell$.

{\em Stage 2}: Define $\delta^{(k)} = \sum_{j=1}^{n} m_j - \sum_{j=1}^{n} 
{m}_{j}^{(k)}$. It is clear that $\delta^{(k)} = \delta^{(0)}$, for 
$k=1,2, \cdots, l$.  In the Schur-Horn situation $\delta^{(0)}$ 
vanishes 
by hypothesis. We will now employ two-mode squeezing transformations 
$S_\mu$ to
rectify this `departure' from the Schur-Horn situation.

We know that $\delta^{(\ell)}=\delta^{(0)}$ is nonnegative. Assume
$\delta^{(0)} > 0$ ( if $\delta^{(0)}=0$, one will move directly to 
Stage~4, 
as will become evident below). Define $\epsilon_{\ell+1} = 
m_{\ell+1}-{m}_{\ell+1}^{(\ell)}$.
Assume $\delta^{(\ell)} \geq 2 \epsilon_{\ell+1}$ ( if this is not the 
case 
one will
 move to Stage~3). Carry out a two-mode squeezing transformation 
$S_\mu$ between 
the ${(\ell+1)}^{\rm th}$ and  ${n}^{\rm th}$ modes, raising the 
corresponding diagonal entries ${m}^{(\ell)}_{\ell+1}$, 
${m}_{n}^{(\ell)}={m}_{n}^{(0)}=
\kappa_n $ by equal magnitude to $m_{\ell+1}$, ${m}_{n}^{(\ell)}+ 
\epsilon_{\ell+1}$
with no change in the other diagonal entries, so that
\begin{eqnarray}
m^{(\ell+1)} &=& (m_1, \cdots , m_{\ell+1}, {m}_{\ell+2}^{(\ell+1)}, 
\cdots , {m}_{n}^{(\ell+1)}),\nonumber\\  
{m}_{j}^{(\ell+1)} &=& {m}_{j}^{(\ell)}, \;\;\; \forall \;\ell+2 \leq j 
\leq n-1,  \nonumber \\
{m}_{n}^{(\ell+1)}&=& {m}_{n}^{(\ell)} + \epsilon_{\ell+1}=
\kappa_n + \epsilon_{\ell+1}.
\end{eqnarray}  
We can now repeat this kind of two-mode squeezing transformation between 
the ${(\ell+2)}^{\rm th}$ mode and the $n^{\rm th}$ mode, and so on. 
Assume we are
able to carry out this process only $r$ times. We will have, at the end of 
it, 
\begin{eqnarray}
m^{(\ell+r)}&=&(m_1,  \cdots, m_{\ell+r}, {m}_{\ell+r+1}^{(\ell+r)}, 
\cdots, {m}_{n}^{(\ell+r)}),\nonumber \\
{m}_{j}^{(\ell+r)} &=& {m}_{j}^{(\ell)}, \;\;\; \forall \;\ell+r+1 \leq j 
\leq n-1,  \nonumber \\
{m}_{n}^{(\ell+r)} &=& \kappa_n + \epsilon_{\ell+1} + \epsilon_{\ell+2}+ 
\cdots +\epsilon_{\ell+r},   
\end{eqnarray}
so that ${\delta}^{(\ell+r)}= \delta^{(0)} -2(\epsilon_{\ell+1} + 
\epsilon_{\ell+2}+ 
\cdots +\epsilon_{\ell+r})$. Clearly, $0 \leq \delta^{(\ell+r)} < 2 
\epsilon_{\ell+r+1} =
2(m_{\ell+r+1} - {m}_{\ell+r+1}^{(\ell+r)})$ (the last inequality 
encodes the fact that we could not 
carry out the Stage~2 operation one more time).

{\em Stage 3}: Assume $\delta^{(\ell+r)} > 0$ ( if $\delta^{(\ell+r)}=0$, 
we move
directly to Stage~4). Carry out a two-mode canonical transformation 
between
the ${(\ell+r+1)}^{\rm th}$ mode and the $n^{\rm th}$ mode, taking the 
corresponding diagonal entries ${m}_{\ell+r+1}^{(\ell+r)}$, 
${m}_{n}^{(\ell+r)}$ 
to
$m_{\ell+r+1}={m}_{\ell+r+1}^{(\ell+r)} + \epsilon_{r+\ell+1}$ and 
${m}_{n}^{(\ell+r+1)}=
{m}_{n}^{(\ell+r)} + \delta^{(\ell+r)} - \epsilon_{r+\ell+1}$ 
respectively, 
leaving
the other diagonals invariant, so that we have
\begin{eqnarray}
{m}^{(\ell+r+1)} &=& (m_1,  \cdots, m_{\ell+r+1}, 
{m}^{(\ell+r+1)}_{\ell+r+2}, 
\cdots ,{m}^{(\ell+r+1)}_{n}),\nonumber \\
{m}_{j}^{(\ell+r+1)}&=& {m}_{j}^{\ell}
< m_j , \;\;\; \forall \; \ell+r+2 \leq j \leq n-1, 
 \nonumber \\
&&\sum_{j=\ell+r+2}^{n} {m}_{j}^{(\ell+r+1)} = \sum_{j=\ell+r+2}^{n} m_j .
\end{eqnarray}  
i.e., the situation in respect of the remaining 
$n-(\ell+r+1)$ (or $n-\ell-r$ if $\delta^{(\ell +r)} =0$) 
modes is  precisely
of the Schur-Horn type, suggesting that we 
 deploy the beam splitter transformation 
$\,n-l-r-2\,$  (or $\,n-l-r-1\,$) times. 

{\em Stage 4} : Note that at the end of Stage~3 we 
have ${m}_{n}^{(\ell+r+1)}$ 
larger than $m_n$ precisely by the sum of the amounts by which 
${m}_{\ell+r+1+j}^{(\ell+r+1)}$, for 
$1 \leq j \leq n-\ell-r-2$, are less than $m_{\ell+r+1+j}$.  
 Therefore, for each value of $j$ in this 
range, we effect a beam splitter 
transformation
connecting the ${(\ell+r+1+j)}^{\rm th}$ mode to the $n^{\rm th}$ mode,
raising ${m}^{(\ell+r+1)}_{\ell+r+1+j}$ to $m_j$ and correspondingly 
pulling
${m}_{n}^{(\ell+r+1+j)}$ down by an equal amount. It is clear that at 
the 
end of these $n-\ell-r-2$  (or $n-\ell-r-1$) redistributions, the 
diagonals will be
precisely $m$. That is, ${m}^{(n-1)}=m$. This completes  proof of the
theorem. 

We have taken maximal advantage of the simpler two-mode 
transformations $S_\theta,\,S_\mu$. The former was deployed 
 $r$ times in Stage~1 and $n-\ell-r-2$  (or $n-\ell-r-1$) times 
in Stage~4, and the latter $\ell$ times in Stage~2.  The more 
general two-mode transformation was deployed (at the most) once in 
Stage~3. 
  
As illustration, and for comparison with  
Ref.\,\cite{eisert}, 
we apply our procedure 
to the example noted  
  after the  statement of the theorem. The difference between  
$\sum_{j=1}^{7} {m}_{j} = 63$ and $\sum_{j=1}^{7} \kappa_j =45$  
 indicates the amount of squeezing that will have to be deployed 
at Stages 3 and 4. We have   
$m^{(0)} \equiv \kappa = (1,2,3,4,5,12,18)$; 
$m^{(1)} = (6,2,3,4,5,7,18)$; 
$m^{(2)} = (6,7,3,4,5,2,18)$; 
$m^{(3)} = (6,7,8,4,5,2,23)$; 
$m^{(4)} = (6,7,8,9,5,2,26)$; 
$m^{(5)} = (6,7,8,9,10,2,21)$; 
and $m^{(6)} = (6,7,8,9,10,11,12) = m$. 
The number of two-mode transformations required at the four stages are
2, 1, 1, and 2 respectively.
 Note that $m^{(k)}$ dominates $m^{(k+1)}$, for $k = 0,1, \cdots, 5$.

\end{document}